\documentclass[pdftex,cite,aps,prl,superscriptaddress,amsmath,amsfonts,amssymb,twocolumn,showpacs]{revtex4}
\usepackage{wasysym}
\renewcommand{\vec}[1]{\mathbf{#1}} \usepackage{graphics}
\newcommand{\dS}{d\vec{S}}
\usepackage[pdftex]{graphicx}

\newcommand{\figref}[1]{Fig.~\ref{fig:#1}}

\newcommand{\figreftwo}[2]{Figs.~\ref{fig:#1} and~\ref{fig:#2}}

\renewcommand{\eqref}[1]{Eq.~\ref{eq:#1}}

\newcommand{\citeasnoun}[1]{Ref.~\onlinecite{#1}}

\begin{document}

\title{Computation and visualization of Casimir forces in arbitrary
geometries: non-monotonic lateral-wall forces and failure of
proximity-force approximations}

\author{Alejandro Rodriguez}
\affiliation{Center for Materials Science and Engineering,
Massachusetts Institute of Technology, Cambridge, MA 02139}
\author{Mihai Ibanescu}
\affiliation{Center for Materials Science and Engineering,
Massachusetts Institute of Technology, Cambridge, MA 02139}
\author{Davide Iannuzzi}
\affiliation{Faculty of Sciences, Department of Physics and Astronomy, Vrije Universiteit Amsterdam, The Netherlands}
\author{Federico Capasso}
\affiliation{Department of Physics, Harvard University, Cambridge, MA
02139}
\author{J. D. Joannopoulos}
\affiliation{Center for Materials Science and Engineering,
Massachusetts Institute of Technology, Cambridge, MA 02139}
\author{Steven G. Johnson}
\affiliation{Center for Materials Science and Engineering,
Massachusetts Institute of Technology, Cambridge, MA 02139}

\begin{abstract}
  We present a method of computing Casimir forces for arbitrary
  geometries, with any desired accuracy, that can directly exploit the
  efficiency of standard numerical-electromagnetism techniques.  Using
  the simplest possible finite-difference implementation of this
  approach, we obtain both agreement with past results for
  cylinder-plate geometries, and also present results for new
  geometries.  In particular, we examine a piston-like problem
  involving two dielectric and metallic squares sliding between two
  metallic walls, in two and three dimensions, respectively, and
  demonstrate non-additive and non-monotonic changes in the force
  due to these lateral walls.
\end{abstract}

\maketitle 

Casimir forces arise between macroscopic objects due to changes in the
zero-point energy associated with quantum fluctuations of the
electromagnetic field~\cite{Lifshitz80}. This spectacular effect has
been subject to many experimental validations, as reviewed in
\citeasnoun{Onofrio06}. All of the experiments reported so far have
been based on simple geometries (parallel plates, crossed cylinders,
or spheres and plates). For more complex geometries, calculations
become extremely cumbersome and often require drastic approximations,
a limitation that has hampered experimental and theoretical work
beyond the standard geometries.

\begin{figure}[hbt]
\includegraphics[width=0.45\textwidth]{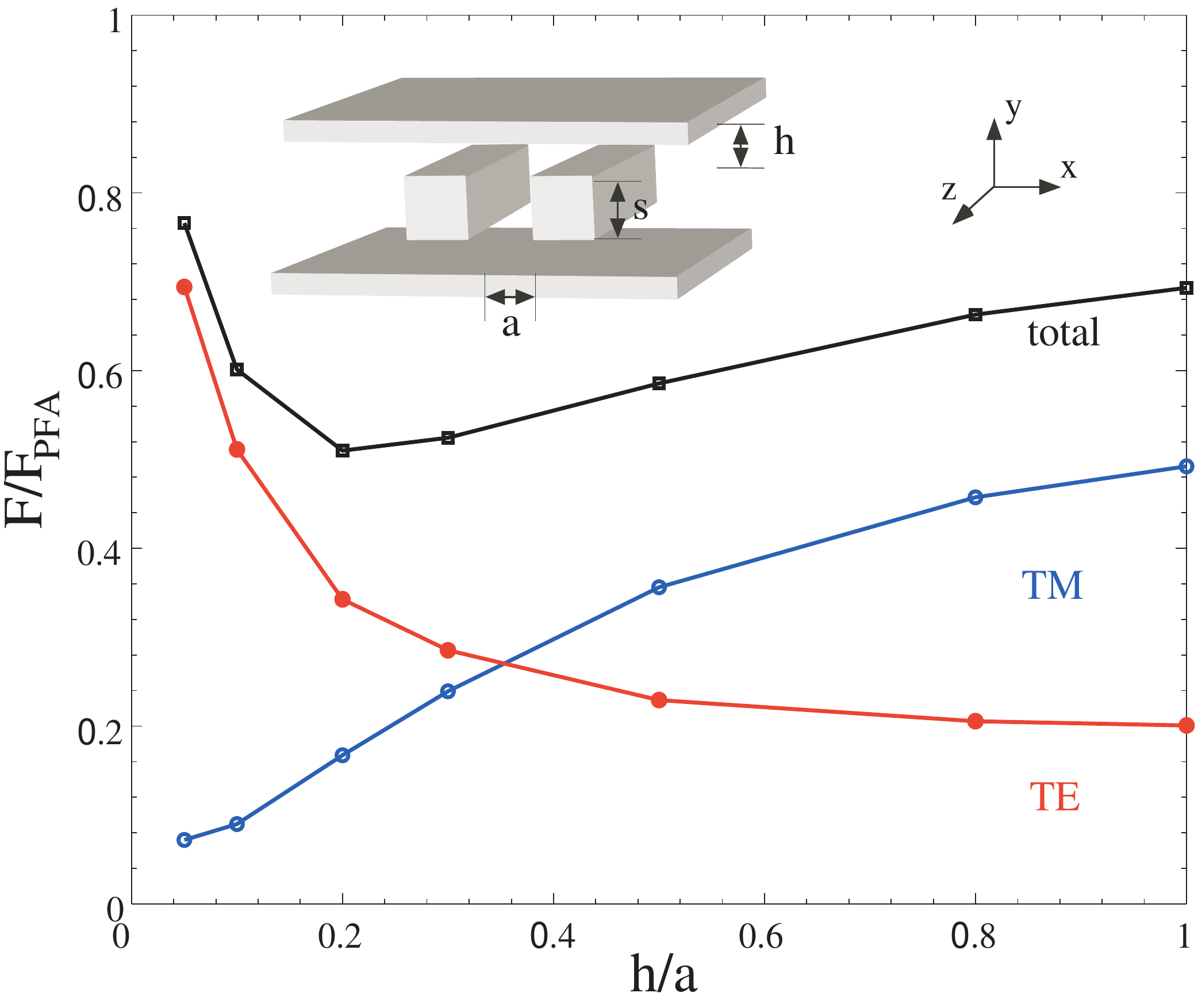}
\caption{Casimir force per unit length between metal squares
$F/F_\textrm{PFA}$, vs. distance from metal plate $h$ (inset),
normalized by the total TE+TM force per unit length obtained using
the PFA, $F_\textrm{PFA}=~\hbar cs \zeta(3)/480\pi a^4$. The total force
is plotted (black squares) along with the TE (red dots) and TM (blue
circles) contributions.}
\label{fig:3d-metal}
\end{figure}

In this letter, we present a method to compute Casimir forces in
arbitrary geometries and materials, with no uncontrolled
approximations, that can exploit the efficient solution of
well-studied problems in classical computational
electromagnetism. Using this method, which we first test for
geometries with known solutions, we predict a non-monotonic
change in the force arising from lateral side walls in a less-familiar
piston-like geometry (\figref{3d-metal}). Such a lateral-wall force cannot
be predicted by ``additive'' methods based on proximity-force or other
purely two-body--interaction approximations, due to symmetry, and it
is difficult to find a simple correction to give a non-monotonic
force.  We are able to compute forces for both perfect metals and
arbitrary dispersive dielectrics, and we also obtain a visual map of
the stress tensor that directly depicts the interaction forces between
objects.

The Casimir force was originally predicted for parallel metal plates,
and the theory was subsequently extended to straighforward formulas
for any planar-multilayer dielectric distribution
$\varepsilon(x,\omega)$ via the generalized Lifshitz
formula~\cite{Tomas02}. In order to handle more arbitrary geometries,
two avenues have been pursued. First, one can employ approximations
derived from limits such as that of parallel plates; these methods
include the proximity-force approximation (PFA) and its
refinements~\cite{Bordag06}, renormalized
Casimir-Polder~\cite{Tajmar04} or semi-classical
interactions~\cite{Schaden98}, multiple-scattering
expansions~\cite{Balian78}, classical ray optics~\cite{Jaffe04}, and
various perturbative techniques~\cite{emig01,Rodrigues06}. Such
methods, however, involve uncontrolled approximations when applied to
arbitrary geometries outside their range of applicability, and have
even been observed to give qualitatively incorrect
results~\cite{Kenneth02,gies06:edge}.  Therefore, researchers have
instead sought numerical methods applicable to arbitrary geometries
that converge to the exact result given sufficient computational
resources. One such method uses a path-integral representation for the
effective action~\cite{emig05}, and has predicted the force between a
cylinder and a plane or between corrugated surfaces.
\citeasnoun{emig05} uses a surface parameterization of the fields
coupled via vacuum Green's functions, requiring $O(N^2)$ storage and
$O(N^3)$ time for $N$ degrees of freedom, making scaling to three
dimensions (3d) problematic. Another exact method is the ``world-line
approach''~\cite{gies03, gies06:PFA,gies06:edge}, based on Monte-Carlo
path-integral calculations. (The scaling of the world-line method
involves a statistical analysis, determined by the relative feature
sizes in the geometry, that is beyond the scope of this Letter.)
Furthermore, the methods of \citeasnoun{emig05} and
\citeasnoun{gies03} have currently only been demonstrated for
perfect-metallic $z$-invariant structures---in this case, the vector
unknowns can be decomposed into TE ($\vec{E} \cdot \hat{\vec{z}} = 0$)
and TM ($\vec{H} \cdot \hat{\vec{z}} = 0$) scalar fields with
Dirichlet or Neumann boundary conditions---although generalizations
have been proposed~\cite{emig04_1}.  Here, we propose a method based
on evaluation of the mean stress tensor via the
fluctuation-dissipation theorem, which only involves repeated
evaluation of the electromagnetic imaginary-frequency Green's
function.  For a volume discretization with $N$ degrees of freedom and
an efficient iterative solver, this requires $O(N)$ storage and
$O(N^{2-1/d})$ time in $d$ dimensions.  Furthermore, because
evaluation of the Green's function is such a standard problem in
classical computational electromagnetism, it will be possible to
exploit many developments in fast solvers, such as finite-element, or
boundary-element methods~\cite{chew01}.  To illustrate the method, our
initial implementation is based on the simplest-possible
finite-difference frequency-domain (FDFD) method, as described below.

As derived by Dzyaloshinski\u{\i} \textit{et al.}~\cite{Lifshitz80},
the net Casimir force on a body can be expressed as an integral over
any closed surface around the body of the mean electromagnetic stress
tensor $\langle T_{ij} \rangle$, integrated over imaginary
frequencies $\omega=iw$:
\begin{equation}
F_i = \int_0^\infty \frac{\hbar dw}{\pi} \oiint_{\mathrm{surface}} \sum_j \langle T_{ij}(\vec{r},iw) \rangle \, dS_j \, .
\label{eq:F}
\end{equation}
For a 3d $z$-invariant structure, the $z$ integral is replaced by an
integral over the corresponding wavevector, resulting in a net force
per unit length.  The stress tensor is defined as usual by:
\begin{multline}
\left\langle T_{ij} (\vec{r},iw) \right\rangle = \left\langle
H_{i}(\vec{r})\,H_{j}(\vec{r})\right\rangle
-\frac{1}{2}\delta_{ij}\sum_k\left\langle
H_{k}(\vec{r})\,H_{k}(\vec{r})\right\rangle \\ +
\varepsilon(\vec{r},iw) \left[ \left\langle
E_{i}(\vec{r})\,E_{j}(\vec{r})\right\rangle -\frac{1}{2}\delta_{ij}\sum_k
\left\langle E_k(\vec{r})\,E_k(\vec{r})\right\rangle \right] \,.
\end{multline}
The connection to quantum mechanics arises from the correlation
functions of the fluctuating fields, such as $\langle E_i E_j
\rangle$, given via the fluctuation-dissipation theorem in terms of
the imaginary-$\omega$ Green's function $G_{ij}(iw;\vec{r}
-\vec{r'})$:
\begin{gather}
\left\langle E_{i}(\vec{r}) E_{j}(\vec{r}')\right\rangle 
    = \frac{w^2}{c^2} G_{ij}(iw;\vec{r} - \vec{r}') \\
\left\langle H_{i}(\vec{r}) H_{j}(\vec{r}')\right\rangle 
    =-(\nabla\times)_{i\ell}(\nabla'\times)_{jm} G_{\ell m}(iw;\vec{r} - \vec{r}') \, ,
\end{gather}
where the Green's function $G_{ij}$ solves the equation:
\begin{equation}
\left[\nabla\times\nabla\times+\frac{w^2}{c^2}\varepsilon(\vec{r},iw)\right]\vec{G}_{j}(iw;\vec{r}-\vec{r}')=\hat{\vec{e}}_{j}\delta(\vec{r}-\vec{r}') 
\label{eq:Green-re}
\end{equation}
for a unit vector $\hat{\vec{e}}_{j}$ in the $j$ direction, and obeys
the usual boundary conditions on the electric field from classical
electromagnetism. (The above expressions are at zero temperature;
the nonzero-temperature force is found by changing $\int dw$ in
\eqref{F} into a discrete summation~\cite{Lifshitz80}.)  Although the
Green's function (and thus $T_{ij}$) is formally infinite at
$\vec{r} = \vec{r}'$, this divergence is conventionally removed by
subtracting the vacuum Green's function; in a numerical method with
discretized space, as below, there is no divergence and no additional
regularization is required.  (The vacuum Green's function gives zero
net contribution to the $\dS$ integral, and therefore need not be
removed as long as the integrand is finite.)

Historically, this stress-tensor expression was used to derive the
standard Lifshitz formula for parallel plates, where $G_{ij}$ is
known analytically.  However, it also forms an ideal starting point
for a computational method, because the Green's function for arbitrary
geometries is routinely computed numerically by a variety of
techniques~\cite{chew01}.  Furthermore, the problem actually becomes
easier for an imaginary $\omega$.  First, for an imaginary $\omega$,
the linear operator in \eqref{Green-re} is real-symmetric and
positive-definite for $w \neq 0$, since the dielectric function
$\varepsilon(\omega)$ is purely real and positive along the
imaginary-$\omega$ axis for physical materials without gain, due to
causality.  Second, the imaginary-$\omega$ Green's function is
exponentially decaying rather than oscillating, leading to a
well-behaved non-oscillatory integrand in \eqref{F}.

To illustrate this method, we employed the simplest possible
computational technique: we perform a FDFD discretization of
\eqref{Green-re} with a staggered Yee grid~\cite{Christ87} and
periodic boundaries, inverting the linear operator by a
conjugate-gradient method.  The presence of discontinuous material
interfaces degrades second-order finite-difference methods to only
first-order accuracy, and the uniform spatial resolution is also
suboptimal, but we found FDFD to be nevertheless adequate for small 2d
geometries.  The periodicity leads to artificial ``wrap-around''
forces that decay rapidly with cell size $L$ (at least as $1/L^3$ in
2d and $1/L^4$ in 3d); we chose cell sizes large enough to make these
contributions negligible ($< 1\%$).

The computational process is as follows: pick some surface/contour
around a given body, evaluate the Green's function for every grid
point on this surface in order to compute the surface integral of the
stress tensor, which is then integrated over $w$ by adaptive
quadrature.

\begin{figure}[ht]
\includegraphics[width=0.45\textwidth]{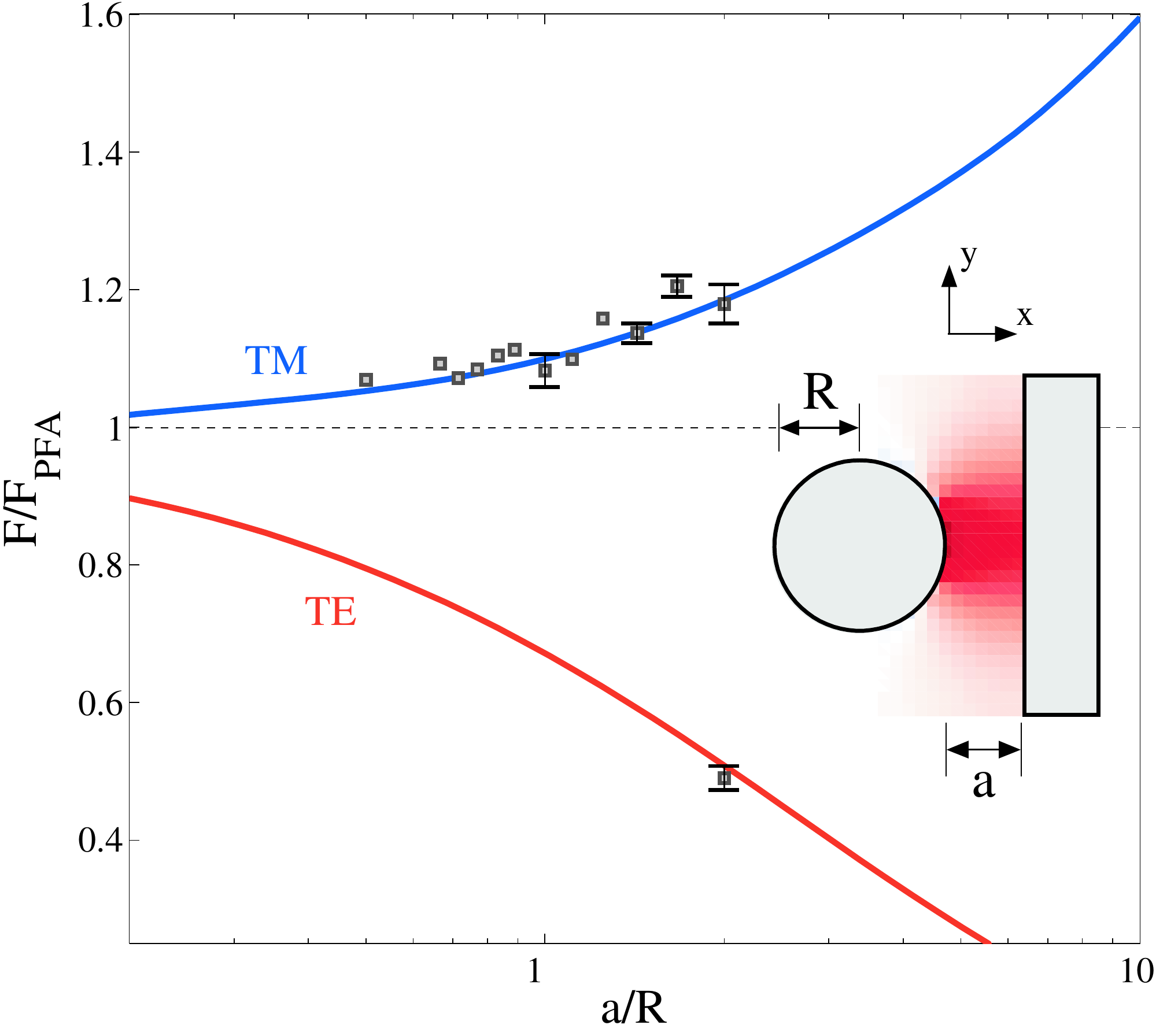}
\caption{Casimir force between a radius-$R$ cylinder and a plate
(inset), relative to the proximity-force approximation
$F_\mathrm{PFA}$, vs. normalized separation $a/R$. The solid lines are
the Casimir force computed in \citeasnoun{emig06} for TE (gray) and TM
(blue) polarizations, along with results computed by our method with a
simple finite-difference discretization (gray squares). Error bars were
estimated for some data points by using computations at multiple
spatial resolutions.  Inset shows interaction stress tensor $\Delta
\langle T_{xx} \rangle$ at a typical imaginary frequency $w =
2\pi c/a$, where red indicates attractive stress.}
\label{fig:cylinder}
\end{figure}

Before we attempt to study new geometries with our method, it is
important to check it against known results.  The simplest cases, of
parallel metallic or dielectric plates, of course match the known result from
the Liftshitz formula and are not reproduced here.  A more complicated
geometry, consisting of a perfect metallic cylinder adjacent to a
perfect metallic plate in 3d, was solved numerically by
\citeasnoun{emig06}, to which our results are compared in
\figref{cylinder}.  \citeasnoun{emig06} used a specialized
Fourier-Bessel basis specific to this cylindrical geometry, which
should have exponential (spectral) convergence. Our use of a simple
uniform grid was necessarily much less efficient, especially with the
first-order accuracy, but was able to match the \citeasnoun{emig06}
results within $ \sim 3\%$ using reasonable computational resources. A
simple grid has the advantage of being very general, as illustrated
below, but other general bases with much greater efficiency are
possible using finite-element or boundary-element methods; the latter,
in particular, could use a spectral Fourier basis analogous to
\citeasnoun{emig06} and exploit a fast-multipole or similar $O(N \log
N)$ solver technique~\cite{chew01}.

Also shown, in the inset of \figref{cylinder}, is a plot of the
interaction stress-tensor component $\Delta \langle T_{xx} \rangle$ at
a typical imaginary frequency $w = 2\pi c/a$.  By ``interaction''
stress-tensor $\Delta \langle T_{ij}\rangle$, we mean the total
$\langle T_{ij}\rangle$ of the full geometry minus the sum of the
$\langle T_{ij}\rangle$'s computed for each body in isolation.  Here,
the stress tensors of the isolated cylinder and plate have been
subtracted, giving us a way to visualize the force due to the
interaction.  As described below, such stress plots reveal the
regions in which two objects most strongly affect one another, and
therefore reveal where a change of the geometry would have the most
impact.  (In contrast, \citeasnoun{gies06:edge} plots an
interaction-energy density that does not directly reveal the force,
since the force requires the energy to be differentiated with respect
to $a$.  For example, \citeasnoun{gies06:edge}'s subtracted energy
density apparently goes nearly to zero as a metallic surface is
approached, whereas the stress tensor cannot since the stress
integration surface is arbitrary.)

We now consider a more complicated geometry in which there are
interactions between multiple bodies: a 3d ``piston''-like structure,
shown in the inset of \figref{3d-metal}, consisting of two
$z$-invariant metal $s \times s$ squares separated by a distance $a$
from one another (here, $s = a$) and separated by a distance $h$ from
infinite metal plates on either side.  We then compute the Casimir
force per unit $z$ between the two squares as a function of the
separation $h$.  The result for perfect conductors is shown in
\figref{3d-metal}, plotted for the TE and TM polarizations and also
showing the total force. (Error bars are not shown because the
estimated error is $< 1\%$.)  In the limit of $h\rightarrow 0$, this
structure approaches the ``Casimir piston,'' which has been solved
analytically~\cite{Hertzberg05,Marachevsky07} (and also in 2d for the
TM polarization~\cite{cavalcanti04}). Our results, extrapolated to
$h=0$, agree with these results to within $\approx 2\%$ (although we
have computational difficulties for small $h$ due to the high
resolution required to resolve a small feature in FDFD).  For $h > 0$,
however, the result is surprising in at least two ways.  First, the
total force is \emph{non-monotonic} in $h$, due to a competition
between the TE and TM contributions to the forces.  Second, the $h$
dependence of the force is a \emph{lateral} effect of the parallel
plates on the squares, which would be zero by symmetry in PFA or any
other two-body--interaction approximation. Although lateral-wall
effects can clearly arise qualitatively in various approximations,
such as in ray optics or in PFA restricted to ``line-of-sight''
interactions, non-monotonicity is more surprising\footnote{A
  publication with R.~L.~Jaffe is in preparation that shows
  non-monotonicity arising in ray optics.}.  Also, in the large-$h$
limit, the force remains different from PFA due to finite-$s$ ``edge''
effects~\cite{gies06:edge}, which are captured by our method.

\begin{figure}[t]
\includegraphics[width=0.45\textwidth]{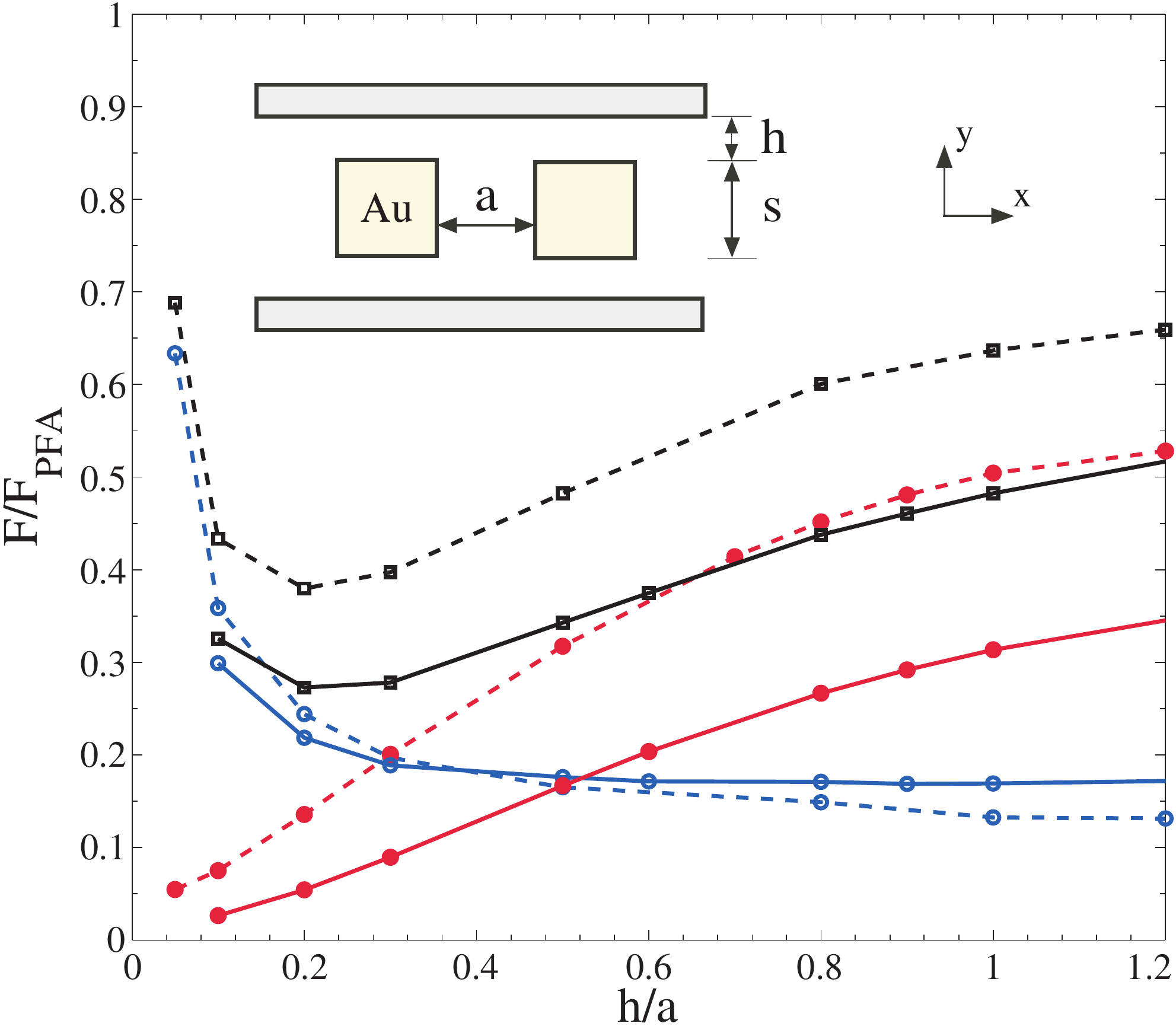}
\caption{\emph{Solid lines:} Casimir force between 2d gold squares
  $F/F_\textrm{PFA}$, vs. distance from metal plate $h$ (inset), using
  experimental $\varepsilon(\omega)$~\cite{Brevik05}, normalized by
  the total force obtained using the PFA. (Here, the PFA force is
  computed for $x$-infinite gold slabs). The total force is plotted
  (black squares) along with the TE (red dots) and TM (blue circles)
  contributions. \emph{Dashed lines:} force for 2d perfect-metal
  squares, normalized by the perfect-metal PFA force $F_{\textrm{PFA}}
  = \hbar cs \zeta(3)/ 8\pi a^3$.}
\label{fig:gold-blocks}
\end{figure}

Our method is also capable, without modification, of handling
dielectric materials.  This is demonstrated in \figref{gold-blocks},
where the Casimir force is shown for the case where the squares are
made of gold, using the experimental Drude $\varepsilon(\omega)$ from
\citeasnoun{Brevik05} for a separation $a=1\mu\mathrm{m}$.  Here, our
calculation is for a purely 2d geometry (equivalently, 3d restricted
to $z$-invariant fields/currents), and for comparison we also plot the
corresponding 2d force for perfect-metal squares (although the two
cases are normalized differently as described in the caption).  As
might be expected, the dielectric squares have a weaker interaction
than the perfect-metal squares, but are still non-monotonic.  Note
also the qualitative similarity between the perfect-metal results of
\figreftwo{3d-metal}{gold-blocks}, reflecting the fact that the force
contributions are dominated by the zero-wavevector ($z$-invariant)
fields.  Extrapolated to $h=0$, the perfect-metal TM force agrees with
the known analytical result~\cite{cavalcanti04} to within $\approx
3\%$.

\begin{figure}[ht]
\includegraphics[width=0.5\textwidth]{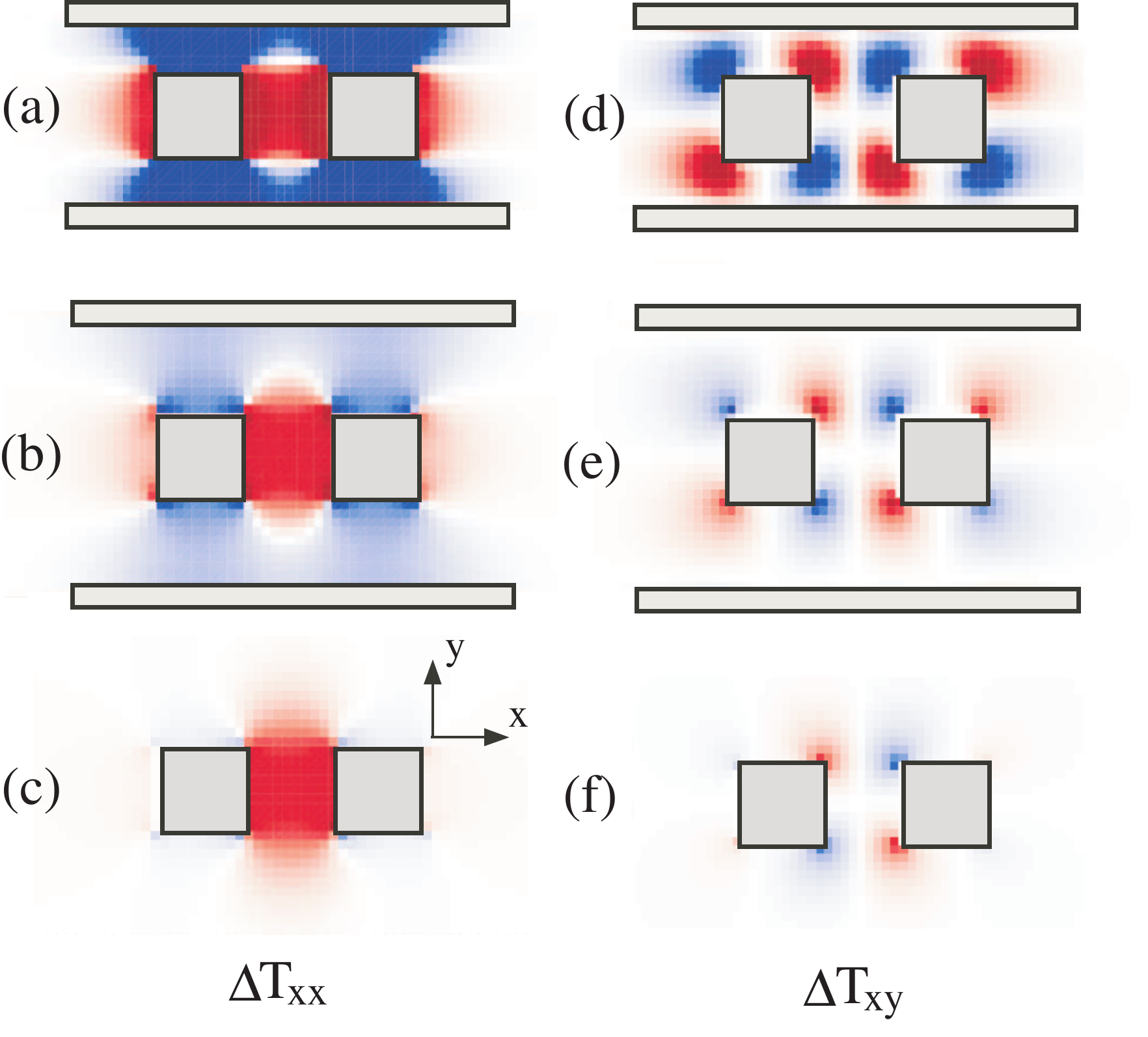}
\caption{(a--f): TM stress map of the 2d-analogous geometry of
\figref{3d-metal} for various $h$.  The intearaction stress tensors
$\langle T_{xx} \rangle$ (left) and $\langle T_{xy}
\rangle$ (right) for: (a),(d): $h = 0.5a$; (b),(e): $h = a$; and
(c),(f): $h = 2a$, where blue/white/red = repulsive/zero/attractive.}
\label{fig:sub}
\end{figure}

To further explore the source of the $h$-dependence, we plot the TM
interaction-stress maps $\Delta \langle T_{xx} \rangle$ and $\Delta
\langle T_{xy} \rangle$ in \figref{sub}, for the 2d perfect-metal
squares from \figref{gold-blocks}. The stress plots of \figref{sub}
are computed at a typical frequency $w = 2\pi c / a$, and for varying
distances from the metal plates ($h = 0.5$, $1.0$, $2.0$). As shown,
the magnitudes of both the $xx$ (a--c) and $xy$ (d--f) components of
the stress tensor change dramatically as the metal plates are brought
closer to the squares. For example, one change in the force integral
comes from $T_{xy}$, which for isolated squares has an asymmetric
pattern at the four corners that will contribute to the attractive
force, whereas the presence of the plates induces a more symmetric
pattern of stresses at the four corners that will have nearly zero
integral.  This results in a decreasing TM force with decreasing $h$.
Because stress maps indicate where bodies interact and with what
signs, it may be useful in future work to explore whether they can be
used to design unusual behaviors such as non-additive, non-monotonic,
or even repulsive forces.

This work was supported in part by the Nanoscale Science and
Engineering Center (NSEC) under NSF contract PHY-0117795, by the
Materials Research Science and Engineering Center program of the NSF
under award DMR-9400334, and by a DOE Comp. Science Grad. Fellowship
under grant DE--FG02-97ER25308. D.~I.~gratefully acknowledges support
from the Netherlands Organisation for Scientific Research (NWO), under
the IRI Scheme \textit{Vernieuwingsimpuls} VIDI-680-47-209.


\begin{thebibliography}{23}
\expandafter\ifx\csname natexlab\endcsname\relax\def\natexlab#1{#1}\fi
\expandafter\ifx\csname bibnamefont\endcsname\relax
  \def\bibnamefont#1{#1}\fi
\expandafter\ifx\csname bibfnamefont\endcsname\relax
  \def\bibfnamefont#1{#1}\fi
\expandafter\ifx\csname citenamefont\endcsname\relax
  \def\citenamefont#1{#1}\fi
\expandafter\ifx\csname url\endcsname\relax
  \def\url#1{\texttt{#1}}\fi
\expandafter\ifx\csname urlprefix\endcsname\relax\def\urlprefix{URL }\fi
\providecommand{\bibinfo}[2]{#2}
\providecommand{\eprint}[2][]{\url{#2}}

\bibitem[{\citenamefont{Lifshitz and Pitaevskii}(1980)}]{Lifshitz80}
\bibinfo{author}{\bibfnamefont{E.~M.} \bibnamefont{Lifshitz}} \bibnamefont{and}
  \bibinfo{author}{\bibfnamefont{L.~P.} \bibnamefont{Pitaevskii}},
  \emph{\bibinfo{title}{Statistical Physics: Part 2}}
  (\bibinfo{publisher}{Pergamon}, \bibinfo{address}{Oxford},
  \bibinfo{year}{1980}).

\bibitem[{\citenamefont{Onofrio}(2006)}]{Onofrio06}
\bibinfo{author}{\bibfnamefont{R.}~\bibnamefont{Onofrio}},
  \bibinfo{journal}{New J.~Phys.} \textbf{\bibinfo{volume}{8}},
  \bibinfo{pages}{237} (\bibinfo{year}{2006}).

\bibitem[{\citenamefont{Toma{\v{s}}}(2002)}]{Tomas02}
\bibinfo{author}{\bibfnamefont{M.~S.} \bibnamefont{Toma{\v{s}}}},
  \bibinfo{journal}{Phys. Rev.~A} \textbf{\bibinfo{volume}{66}},
  \bibinfo{pages}{052103} (\bibinfo{year}{2002}).

\bibitem[{\citenamefont{Bordag}(2006)}]{Bordag06}
\bibinfo{author}{\bibfnamefont{M.}~\bibnamefont{Bordag}},
  \bibinfo{journal}{Phys. Rev.~D} \textbf{\bibinfo{volume}{73}},
  \bibinfo{pages}{125018} (\bibinfo{year}{2006}).

\bibitem[{\citenamefont{Tajmar}(2004)}]{Tajmar04}
\bibinfo{author}{\bibfnamefont{M.}~\bibnamefont{Tajmar}},
  \bibinfo{journal}{Intl. J. Mod. Phys. C} \textbf{\bibinfo{volume}{15}},
  \bibinfo{pages}{1387} (\bibinfo{year}{2004}).

\bibitem[{\citenamefont{Schaden and Spruch}(1998)}]{Schaden98}
\bibinfo{author}{\bibfnamefont{M.}~\bibnamefont{Schaden}} \bibnamefont{and}
  \bibinfo{author}{\bibfnamefont{L.}~\bibnamefont{Spruch}},
  \bibinfo{journal}{Phys. Rev.~A} \textbf{\bibinfo{volume}{58}},
  \bibinfo{pages}{935} (\bibinfo{year}{1998}).

\bibitem[{\citenamefont{Balian and Duplantier}(1978)}]{Balian78}
\bibinfo{author}{\bibfnamefont{R.}~\bibnamefont{Balian}} \bibnamefont{and}
  \bibinfo{author}{\bibfnamefont{B.}~\bibnamefont{Duplantier}},
  \bibinfo{journal}{Ann. Phys.} \textbf{\bibinfo{volume}{112}},
  \bibinfo{pages}{165} (\bibinfo{year}{1978}).

\bibitem[{\citenamefont{Jaffe and Scardicchio}(2004)}]{Jaffe04}
\bibinfo{author}{\bibfnamefont{R.~L.} \bibnamefont{Jaffe}} \bibnamefont{and}
  \bibinfo{author}{\bibfnamefont{A.}~\bibnamefont{Scardicchio}},
  \bibinfo{journal}{Phys. Rev. Lett.} \textbf{\bibinfo{volume}{92}},
  \bibinfo{pages}{070402} (\bibinfo{year}{2004}).

\bibitem[{\citenamefont{Emig et~al.}(2001)\citenamefont{Emig, Hanke,
  Golestanian, and Kardar}}]{emig01}
\bibinfo{author}{\bibfnamefont{T.}~\bibnamefont{Emig}},
  \bibinfo{author}{\bibfnamefont{A.}~\bibnamefont{Hanke}},
  \bibinfo{author}{\bibfnamefont{R.}~\bibnamefont{Golestanian}},
  \bibnamefont{and} \bibinfo{author}{\bibfnamefont{M.}~\bibnamefont{Kardar}},
  \bibinfo{journal}{Phys. Rev. Lett.} \textbf{\bibinfo{volume}{87}},
  \bibinfo{pages}{260402} (\bibinfo{year}{2001}).

\bibitem[{\citenamefont{Rodrigues et~al.}(2006)\citenamefont{Rodrigues, Neto,
  Lambrecht, and Reynaud}}]{Rodrigues06}
\bibinfo{author}{\bibfnamefont{R.~B.} \bibnamefont{Rodrigues}},
  \bibinfo{author}{\bibfnamefont{P.~A.~M.} \bibnamefont{Neto}},
  \bibinfo{author}{\bibfnamefont{A.}~\bibnamefont{Lambrecht}},
  \bibnamefont{and} \bibinfo{author}{\bibfnamefont{S.}~\bibnamefont{Reynaud}},
  \bibinfo{journal}{Phys. Rev. Lett.} \textbf{\bibinfo{volume}{96}},
  \bibinfo{pages}{100402} (\bibinfo{year}{2006}).

\bibitem[{\citenamefont{Kenneth et~al.}(2002)\citenamefont{Kenneth, Klich,
  Mann, and Revzen}}]{Kenneth02}
\bibinfo{author}{\bibfnamefont{O.}~\bibnamefont{Kenneth}},
  \bibinfo{author}{\bibfnamefont{I.}~\bibnamefont{Klich}},
  \bibinfo{author}{\bibfnamefont{A.}~\bibnamefont{Mann}}, \bibnamefont{and}
  \bibinfo{author}{\bibfnamefont{M.}~\bibnamefont{Revzen}},
  \bibinfo{journal}{Phys. Rev. Lett.} \textbf{\bibinfo{volume}{89}},
  \bibinfo{pages}{033001} (\bibinfo{year}{2002}).

\bibitem[{\citenamefont{Gies and
  Klingmuller}(2006{\natexlab{a}})}]{gies06:edge}
\bibinfo{author}{\bibfnamefont{H.}~\bibnamefont{Gies}} \bibnamefont{and}
  \bibinfo{author}{\bibfnamefont{K.}~\bibnamefont{Klingmuller}},
  \bibinfo{journal}{Phys. Rev. Lett.} \textbf{\bibinfo{volume}{97}},
  \bibinfo{pages}{220405} (\bibinfo{year}{2006}{\natexlab{a}}).

\bibitem[{\citenamefont{B{\"{u}}scher and Emig}(2005)}]{emig05}
\bibinfo{author}{\bibfnamefont{R.}~\bibnamefont{B{\"{u}}scher}}
  \bibnamefont{and} \bibinfo{author}{\bibfnamefont{T.}~\bibnamefont{Emig}},
  \bibinfo{journal}{Phys. Rev. Lett.} \textbf{\bibinfo{volume}{94}},
  \bibinfo{pages}{133901} (\bibinfo{year}{2005}).

\bibitem[{\citenamefont{Gies et~al.}(2003)\citenamefont{Gies, Langfeld, and
  Moyaerts}}]{gies03}
\bibinfo{author}{\bibfnamefont{H.}~\bibnamefont{Gies}},
  \bibinfo{author}{\bibfnamefont{K.}~\bibnamefont{Langfeld}}, \bibnamefont{and}
  \bibinfo{author}{\bibfnamefont{L.}~\bibnamefont{Moyaerts}},
  \bibinfo{journal}{J. High Energy Phys.} p. \bibinfo{pages}{018}
  (\bibinfo{year}{2003}).

\bibitem[{\citenamefont{Gies and Klingmuller}(2006{\natexlab{b}})}]{gies06:PFA}
\bibinfo{author}{\bibfnamefont{H.}~\bibnamefont{Gies}} \bibnamefont{and}
  \bibinfo{author}{\bibfnamefont{K.}~\bibnamefont{Klingmuller}},
  \bibinfo{journal}{Phys. Rev. Lett.} \textbf{\bibinfo{volume}{96}},
  \bibinfo{pages}{220401} (\bibinfo{year}{2006}{\natexlab{b}}).

\bibitem[{\citenamefont{Emig and B{\"{u}}scher}(2004)}]{emig04_1}
\bibinfo{author}{\bibfnamefont{T.}~\bibnamefont{Emig}} \bibnamefont{and}
  \bibinfo{author}{\bibfnamefont{R.}~\bibnamefont{B{\"{u}}scher}},
  \bibinfo{journal}{Nucl. Phys. B} \textbf{\bibinfo{volume}{696}},
  \bibinfo{pages}{468} (\bibinfo{year}{2004}).

\bibitem[{\citenamefont{Chew et~al.}(2001)\citenamefont{Chew, Jian-Ming,
  Michielssen, and Jiming}}]{chew01}
\bibinfo{author}{\bibfnamefont{W.~C.} \bibnamefont{Chew}},
  \bibinfo{author}{\bibfnamefont{J.}~\bibnamefont{Jian-Ming}},
  \bibinfo{author}{\bibfnamefont{E.}~\bibnamefont{Michielssen}},
  \bibnamefont{and} \bibinfo{author}{\bibfnamefont{S.}~\bibnamefont{Jiming}},
  \emph{\bibinfo{title}{Fast and Efficient Algorithms in Computational
  Electromagnetics}} (\bibinfo{publisher}{Artech}, \bibinfo{address}{Norwood,
  MA}, \bibinfo{year}{2001}).

\bibitem[{\citenamefont{Christ and Hartnagel}(1987)}]{Christ87}
\bibinfo{author}{\bibfnamefont{A.}~\bibnamefont{Christ}} \bibnamefont{and}
  \bibinfo{author}{\bibfnamefont{H.~L.} \bibnamefont{Hartnagel}},
  \bibinfo{journal}{IEEE Trans. Microwave Theory Tech.}
  \textbf{\bibinfo{volume}{35}}, \bibinfo{pages}{688} (\bibinfo{year}{1987}).

\bibitem[{\citenamefont{Emig et~al.}(2006)\citenamefont{Emig, Jaffe, Kardar,
  and Scardicchio}}]{emig06}
\bibinfo{author}{\bibfnamefont{T.}~\bibnamefont{Emig}},
  \bibinfo{author}{\bibfnamefont{R.~L.} \bibnamefont{Jaffe}},
  \bibinfo{author}{\bibfnamefont{M.}~\bibnamefont{Kardar}}, \bibnamefont{and}
  \bibinfo{author}{\bibfnamefont{A.}~\bibnamefont{Scardicchio}},
  \bibinfo{journal}{Phys. Rev. Lett.} \textbf{\bibinfo{volume}{96}},
  \bibinfo{pages}{080403} (\bibinfo{year}{2006}).

\bibitem[{\citenamefont{Hertzberg et~al.}(2005)\citenamefont{Hertzberg, Jaffe,
  Kardar, and Scardicchio}}]{Hertzberg05}
\bibinfo{author}{\bibfnamefont{M.~P.} \bibnamefont{Hertzberg}},
  \bibinfo{author}{\bibfnamefont{R.~L.} \bibnamefont{Jaffe}},
  \bibinfo{author}{\bibfnamefont{M.}~\bibnamefont{Kardar}}, \bibnamefont{and}
  \bibinfo{author}{\bibfnamefont{A.}~\bibnamefont{Scardicchio}},
  \bibinfo{journal}{Phys. Rev. Lett.} \textbf{\bibinfo{volume}{95}},
  \bibinfo{pages}{250402} (\bibinfo{year}{2005}).

\bibitem[{\citenamefont{Marachevsky}(2007)}]{Marachevsky07}
\bibinfo{author}{\bibfnamefont{V.~M.} \bibnamefont{Marachevsky}},
  \bibinfo{journal}{PRD} \textbf{\bibinfo{volume}{75}}, \bibinfo{pages}{085019}
  (\bibinfo{year}{2007}).

\bibitem[{\citenamefont{Cavalcanti}(2004)}]{cavalcanti04}
\bibinfo{author}{\bibfnamefont{R.~M.} \bibnamefont{Cavalcanti}},
  \bibinfo{journal}{Phys. Rev.~D} \textbf{\bibinfo{volume}{69}},
  \bibinfo{pages}{065015} (\bibinfo{year}{2004}).

\bibitem[{\citenamefont{Brevik et~al.}(2005)\citenamefont{Brevik, Aarseth,
  Hoye, and Milton}}]{Brevik05}
\bibinfo{author}{\bibfnamefont{I.}~\bibnamefont{Brevik}},
  \bibinfo{author}{\bibfnamefont{J.~B.} \bibnamefont{Aarseth}},
  \bibinfo{author}{\bibfnamefont{J.~S.} \bibnamefont{Hoye}}, \bibnamefont{and}
  \bibinfo{author}{\bibfnamefont{K.~A.} \bibnamefont{Milton}},
  \bibinfo{journal}{Phys. Rev.~E} \textbf{\bibinfo{volume}{71}}
  (\bibinfo{year}{2005}).

\end{thebibliography}
\end{document}